\documentclass{jpsj-suppl}
\usepackage{txfonts} 

\def\be{\begin{equation}}
\def\ee{\end{equation}}
\def\bea{\begin{eqnarray}}
\def\eea{\end{eqnarray}}
\def\bear{\begin{array}}
\def\ear{\end{array}}
\def\bfig{\begin{figure}}
\def\efig{\end{figure}}
\def\bcen{\begin{center}}
\def\ecen{\end{center}}
\def\bi{\begin{itemize}}
\def\ei{\end{itemize}}

\def\raw{\rightarrow}

\def\slash{\!\!\! /}

\title{Present Status of Single Pion Production in Neutrino-Nucleus Reactions}

\author{L. \textsc{Alvarez-Ruso}$^{1}$, E. \textsc{Hern\'andez}$^{2}$, J. \textsc{Nieves}$^{1}$ and M. J. \textsc{Vicente Vacas}$^{3}$}

\inst{$^{1}$Instituto de F\'\i sica Corpuscular (IFIC), Centro Mixto
  CSIC-UVEG, E-46071 Valencia, Spain \\
$^{2}$Departamento de F\'\i sica Fundamental 
  e IUFFyM, Universidad de Salamanca, E-37008 Salamanca, Spain\\
$^{3}$Departamento de F\'\i sica Te\'orica and IFIC, Centro Mixto
UVEG-CSIC, E-46071 Valencia, Spain}  

\email{Luis.Alvarez@ific.uv.es}

\recdate{January 31, 2016}

\abst{Some of the recent progress in the physics of pion production induced by neutrinos on nucleons and nuclei is reviewed from a theoretical perspective. The importance of Watson's theorem to reconcile ANL and BNL data with the off-diagonal Goldberger-Treiman relation for the $\Delta(1232)$ is discussed. The disagreement between MiniBooNE data and theoretical calculations is presented in the light of the new MINERvA data. The coherent pion production data on $^{12}$C obtained by MINERvA are also compared to different microscopic and PCAC models.}

\kword{weak interactions, baryon resonances, axial current, PCAC, nuclear effects, final-state interactions}

\begin{document}
\maketitle

\section{Introduction}

A better understanding of weak pion production has been actively pursued in recent years. Single pion production amounts to one of the leading contributions to the inclusive (anti)neutrino-nucleus cross section in energy range of interest for several current and future oscillation experiments. As such, it can be part of the signal or a background that should be precisely constrained. CC$1\pi$ interactions are a source of QE-like events that needs to be subtracted for a proper neutrino energy reconstruction in QE based analyses. NC$1\pi^0$ events in Cherenkov detectors contribute to the $e$-like background in $\nu_e$ appearance measurements. 

Theoretical models have been developed alongside new measurements. The later, predominantly  on carbon targets by MiniBooNE~\cite{AguilarArevalo:2009ww,AguilarArevalo:2010xt,AguilarArevalo:2010bm} and MINERvA~\cite{Eberly:2014mra,Aliaga:2015wva} experiments, have revealed discrepancies with existing theoretical models and among different data sets. The first measurement of NC$\pi^0$ in argon has been recently reported by ArgoNeuT~\cite{Acciarri:2015ncl} albeit with low statistics.  

\section{Pion production on nucleons}

The first requirement for a precise description of neutrino induced pion production on nuclear targets is a realistic model at the nucleon level. Weak pion production on the nucleon is also interesting for hadronic physics as a source of information about axial nucleon-to-resonance transition currents. It is also worth stressing that pion production amplitudes are a key ingredient of the very important 2p2h models of QE-like scattering.

Different theoretical studies have stressed the predominant role of the $\Delta(1232)3/2^+$ in the few-GeV region. The weak nucleon-to-$\Delta(1232)$ transition current can be written in terms of vector and axial form factors (FF), $C_{3-5}^V$ and $C_{3-6}^A$ in the notation of Ref.~\cite{LlewellynSmith:1971zm}. Owing to the symmetry of the conserved vector current under isospin rotations, the vector FF can be cast in terms of the helicity amplitudes extracted in the analysis of pion electroproduction data~\cite{Lalakulich:2005cs,Leitner:2008ue}. Our understanding of the axial part of the transition current
\be
-\mathcal{A}^{\mu }_{N\Delta} = \bar{u}_\alpha(p') \left[\frac{C_3^A}{m_N} (g^{\alpha \mu} q\slash  - q^{\alpha} \gamma^{\mu})+
  \frac{C_4^A}{m_N^2} (g^{\alpha \mu} q\cdot p' - q^{\alpha} {p'}^{\mu})  
  + {C_5^A} g^{\alpha \mu}  + \frac{C_6^A}{m_N^2} q^{\alpha} q^{\mu}\right] \gamma_{5}\, u(p)
\label{NDelta_axial}
\ee
is far more scarce. The most important $N\Delta$ FF is $C^A_5$: the only one that appears at leading order in an expansion of the hadronic tensor in the four-momentum transfer $q^2$
\be
\frac{d\sigma}{dq^2}(q^2=0) \propto \left[C_5^A(0)\right]^2 \,.
\label{eq:q2=0}
\ee
For the  subleading $C_{3,4}^A$ axial FF, the approximations $C_3^A=0, C_4^A=-C_5^A/4$ are often adopted following Adler. As a matter of fact, the available bubble-chamber data on pion production induced by neutrinos on deuterium, taken at ANL and BNL~\cite{Radecky:1981fn,Kitagaki:1986ct} are quite insensitive to their values~\cite{Hernandez:2010bx}.

\subsection{The off-diagonal Goldberger-Treiman relation}

This identity is a direct consequence of PCAC and the pion-pole dominance of $C_6^A$. In the chiral ($m_\pi \raw 0$) limit, $q_\mu \mathcal{A}^{\mu }_{N\Delta}=0$, resulting in
\be
C_5^A(0) =\sqrt{\frac{2}{3}} g_{\Delta N\pi} \,. 
\label{GT}
\ee
The leading axial coupling is related to the $\Delta N \pi$ effective coupling $g_{\Delta N\pi}$ defined as
\be
{\mathcal L}_{\Delta N \pi} = - \frac{g_{\Delta N \pi}}{f_\pi} \bar{\Delta}_\mu  (\partial^\mu \vec{{\pi}}) \vec{T}^\dagger N \,, 
\label{DeltaNpi_vertex}
\ee
which can be extracted from $\pi N$ scattering data. Using the PDG estimate of the $\Delta(1232) \raw N \pi$ decay width to calculate $g_{\Delta N\pi}$, the off-diagonal Goldberger-Treiman (GT) relation [Eq.~(\ref{GT})] gives
\be
C_5^A(0)\big{|}_{\mathrm{GT}} = 1.15 - 1.2 \,.
\label{GTnum}
\ee
Deviations from this GT relation are expected only at the few \% level, as they arise from chiral symmetry breaking. Systematic studies of the corrections to the GT relation using chiral perturbation theory have been reported in Refs.~\cite{Geng:2008bm,Procura:2008ze}.

Fits to ANL and/or BNL data for $\nu_\mu p \to \mu^- p\pi^+$ including only the $\Delta(1232)$ excitation mechanism obtained $C_5^A(0)$ values in agreement with the GT relation~\cite{AlvarezRuso:1998hi,Graczyk:2009qm}. However, in Ref.~\cite{Hernandez:2010bx} $C_5^A$, parametrized as
\be
C_5^A(q^2) = C_5^A(0) \left( 1 - \frac{q^2}{M_{A\Delta}^2} \right)^{-2} \,,
\label{dipole}
\ee
was extracted from ANL and BNL data with low invariant masses ($W_{\pi N} < 1.4$~GeV) in a model that incorporated, besides the $\Delta$ amplitudes, non-resonant contributions at tree level complemented with phenomenological weak FF~\cite{Hernandez:2007qq}; the resulting $C_5^A(0) = 1.00\pm 0.11$ turned out to be 2$\sigma$ below the GT value. Close to threshold, these non-resonant terms are fully determined by the chiral symmetry of strong interactions. Taking them into account is therefore mandatory. The model of Ref.~\cite{Hernandez:2010bx} could be reconciled with the GT relation by simultaneously fitting vector form factors to electron-proton scattering structure function $F_2$~\cite{Graczyk:2014dpa}: $C_5^A(0) = 1.10^{+0.15}_{-0.14}$. Nevertheless, it should realized that the interplay between the real tree-level background and the complex $\Delta$ amplitudes violates Watson's theorem.  

\subsection{The Watson's theorem}

Being a consequence of unitarity and time reversal invariance, it states that in the sum over intermediate states of the $T$-matrix elements
\be
\sum_M\langle M|T|F\rangle^*\langle M|T|I\rangle = -2 {\rm Im}\langle F|T|I\rangle\in\mathbb{R}
\ee
the phases cancel each other to give a real quantity. For the specific case of the $W\, N \raw \pi \, N$ amplitude below the $2 \pi$ production threshold, assuming that $|M\rangle = |F\rangle = |\pi N\rangle$, one finds that 
\be
\langle \pi N|T|\pi N\rangle^* \langle \pi N|T|W N\rangle=-2{\rm Im}\langle \pi N |T| W N\rangle\in \mathbb{R} \,.
\ee
where
\be
\langle \pi N|T|\pi N\rangle  \approx \langle \pi N|T_{\mathrm{strong}}|\pi N\rangle \,.
\ee
That is, the phase of the weak pion production matrix element $\langle \pi N|T|W N\rangle$ is determined by the phase of the strong elastic $\pi N$ amplitude. This derivation is merely schematic because even when the particle content is the same in the intermediate and final states, one still has to consider separately states with different angular momentum and sum over the allowed intermediate states. In Ref.~\cite{Alvarez-Ruso:2015eva}, Watson's theorem was restored in the most relevant $P_{33}$ partial wave by writing
\be
T = T_B + T_\Delta \,e^{i \delta(W,q^2)} 
\ee
and choosing the $\delta(W,q^2)$ phases in such a way that
\be
\left[ \sum_{\rho}(1,1/2,3/2;0,-\rho,-\rho)\,\langle 3/2,M;{0,\rho}|T|0,0;{r,\lambda}\rangle \right] e^{-i \delta_{P_{33}}} \in\mathbb{R} \,.
\label{eq:fase33}
\ee
This can be achieved with two independent phases conveniently parametrized in Ref.~\cite{Alvarez-Ruso:2015eva}. The $P_{33}(W)$ $\pi N$ phase shifts have been taken from the SAID Partial Wave Analysis~\cite{SAID}. This approach, originally introduced by Olsson~\cite{Olsson:1974sw}, has been successfully applied to pion photo and electroproduction~\cite{Carrasco:1989vq,Gil:1997bm}.

A fit to ANL and BNL data ($W_{\pi N} < 1.4$~GeV) with the improved model results in
\be
C_5^A(0) = 1.12 \pm 0.11 \,, \,\,\, M_{A\Delta} = 0.95 \pm 0.06 \,\mathrm{GeV} \,,
\label{fit1}
\ee
in agreement with the GT value of Eq.~(\ref{GTnum}).

The determination of $C_5^A(0)$ suffers from long standing inconsistencies between the ANL and BNL data sets. A recent reanalysis~\cite{Wilkinson:2014yfa} has established that the origin of the discrepancies resides in the flux normalization. New consistent cross sections have then been obtained using flux-normalization independent CC1$\pi$/CCQE ratios and the better understood CCQE cross section in deuterium. Taking advantage of these developments, a new fit has been performed in Ref.~\cite{Alvarez-Ruso:2015eva} using the shape of the original ANL $d\sigma/dq^2$ distribution with $W_{\pi N} < 1.4$~GeV, not affected by the flux normalization uncertainty, and the reanalyzed integrated cross sections for $E_\nu < 1.1$~GeV. This choice intended to minimize the role of heavier resonances, as no invariant mass cut was applied in the new data of Ref.~\cite{Wilkinson:2014yfa}. This fit (see Fig.~\ref{fig:fit2}) led to
\be
C_5^A(0) = 1.14\pm 0.07 \,, \,\,\, M_{A\Delta} = 0.96 \pm 0.07 \,\mathrm{GeV} \,.
\label{fit2}
\ee

The axial coupling $C_5^A(0)$ value obtained with the revisited ANL+BNL data is consistent with the one from the original data but in closer agreement with the GT relation. Without imposing Watson's theorem, the revisited data result in $C_5^A(0)=1.05 \pm 0.07$, clearly below the GT relation. This is in line with the findings of Ref.~\cite{Alam:2015gaa}, where a good description of the revised data is achieved with $C_5^A(0)=1$, using a model with heavier $N^*$ resonances but without unitarity. Remarkably, the consistency between the revised ANL and BNL data sets reduces the error in $C_5^A(0)$ from 10\% to 6\% [compare Eqs.~(\ref{fit1}) and (\ref{fit2})]. On the other hand, the value and error of $M_{A\Delta}$ are practically unaffected by this improvement. This reflects the fact that finer details in the structure of the pion production axial current are harder to pin down from these data. Further insight requires new more precise data from dedicated neutrino experiments on H$_2$/D$_2$ targets or indirectly on multinuclear targets using techniques like the one proposed in Ref.~\cite{Lu:2015hea}.
\bfig[t]
\includegraphics[width=0.5\textwidth]{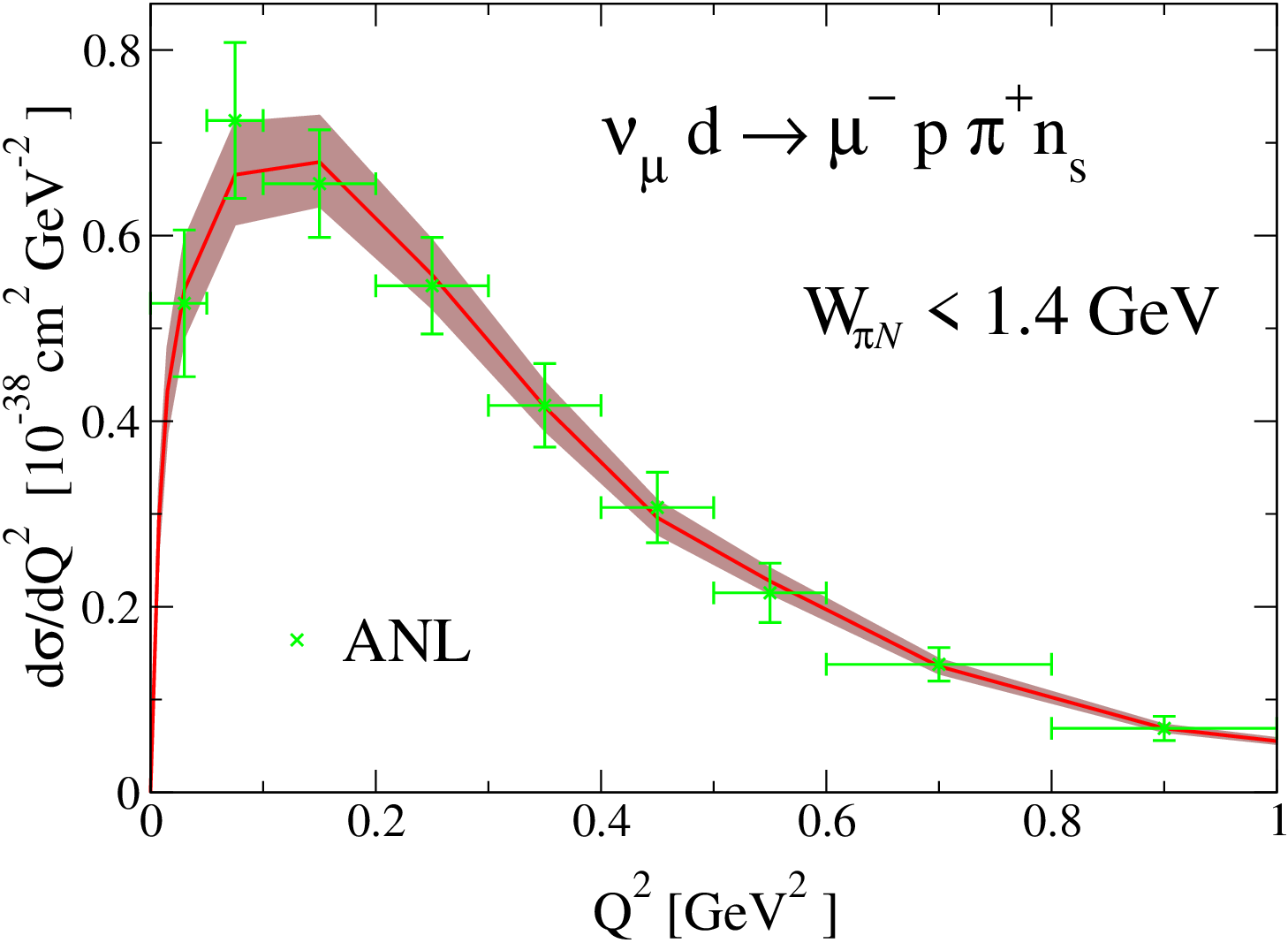}
\includegraphics[width=0.5\textwidth]{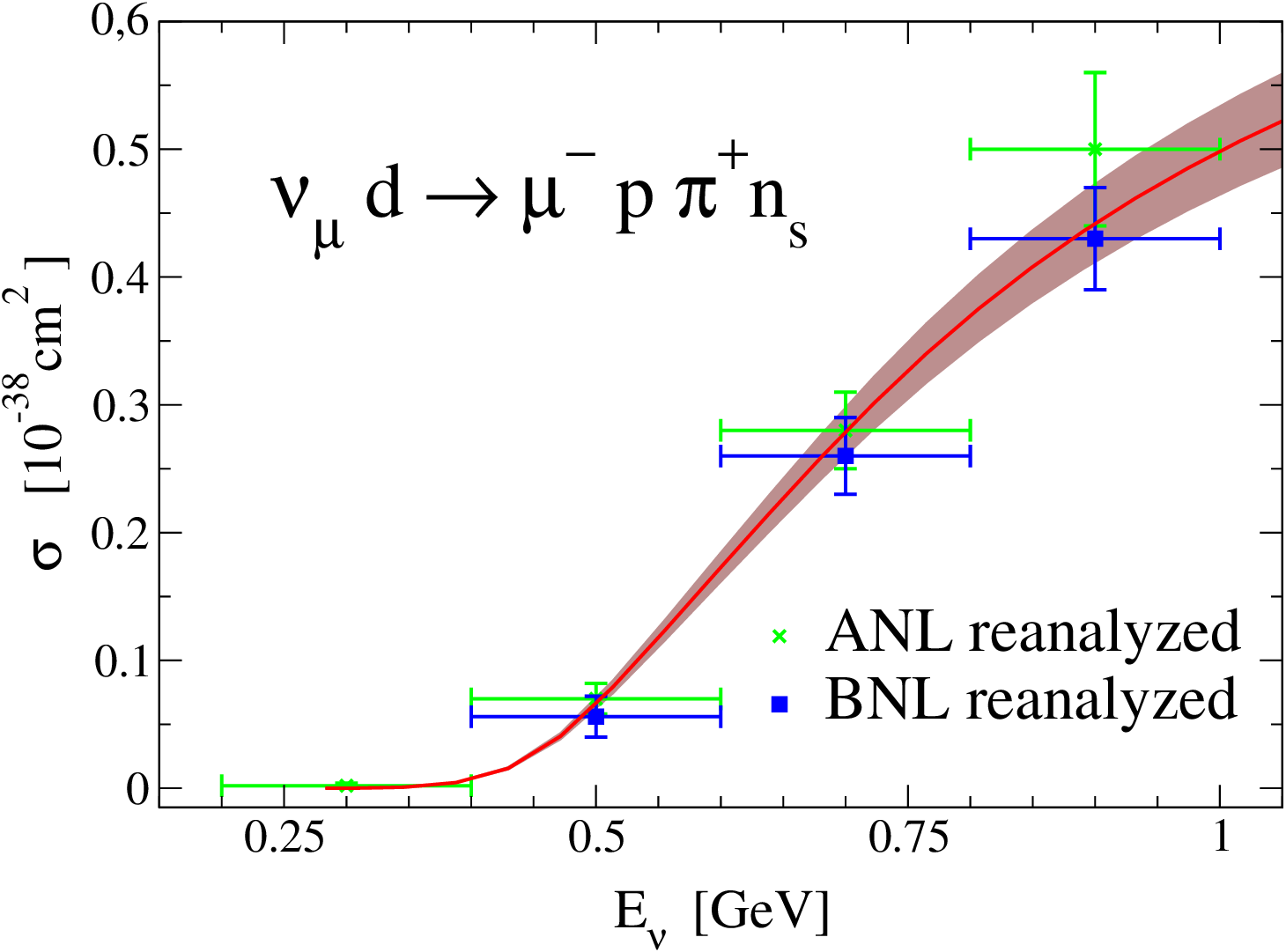}
\caption{Fit~\cite{Alvarez-Ruso:2015eva} to the shape of $d\sigma/dQ^2$ by ANL~\cite{Radecky:1981fn} and the reanalyzed ANL and BNL total cross sections~\cite{Wilkinson:2014yfa} for $\nu_\mu p\to \mu^- p \pi^+$.}
\label{fig:fit2}
\efig

A state of the art description of meson production by means of a dynamical model in coupled channels has been recently extended to the weak reactions~\cite{Nakamura:2015rta}. PCAC is used to derive the axial current. The full amplitudes are the solution of the Lippmann-Schwinger equation. For these reasons, GT relations and the Watson's theorem are respected by construction. The total cross section in the $\nu_\mu p\to \mu^- p \pi^+$ channel is higher than the reanalyzed data of Ref.~\cite{Wilkinson:2014yfa} but one should keep in mind that deuteron corrections have not been considered in Ref.~\cite{Nakamura:2015rta}. Within the spectator approximation, deuteron effects cause a small reduction, not exceeding 8\% even at low $q^2$~\cite{AlvarezRuso:1998hi}, that would make the agreement better. Reference~\cite{Wu:2014rga} obtains an additional reduction at forward angles from the strong interaction of outgoing $p n$ pairs. This calls for a more detailed analysis in the conditions of the ANL and BNL experiments, accounting also for their kinematical cuts. 

\section{Pion production in nuclei}

Modern neutrino experiments are performed on nuclear targets. The presence of the nuclear medium poses additional challenges for the reaction modeling. The initial nucleon is often assumed to be free, with a Fermi momentum according to the global or local Fermi gas models, or interacting with a nuclear mean field. More elaborated descriptions of the initial state like spectral functions~\cite{Benhar:2005dj} and  bound-sate wave functions~\cite{Praet:2008yn} have also become available for baryon resonance excitation and meson production in general. Nevertheless one should stress that at the higher energy transfers present in inelastic processes, the details of nuclear structure are less relevant. The hadronic currents are also modified in the nucleus. Given the prevalent role of the $\Delta(1232)$ excitation in pion production, it is not surprising that the in-medium modification of the $\Delta$ propagator is very important. The main effect is the increase of the $\Delta(1232)$  width (broadening) by many body processes: $\Delta \,N \raw N \, N$, $\Delta \,N \raw N \, N \, \pi$, $\Delta \,N \, N \raw N \, N \, N$. In their way out of the nucleus, pions undergo final state interactions (FSI). They can be absorbed, change their energy, angle and charge. In particular, in CC interactions, there is a considerable side feeding from the dominant $\pi^+$ production to the $\pi^0$ channel~\cite{Leitner:2006ww}. At high momentum transfers, low energy pions can be produced in secondary collisions of nucleons knocked out in QE interactions~\cite{Leitner:2006ww}. The imprint of the strong-interacting environment on the observables is therefore quite significant, obscuring the connection between primary interactions and measured quantities.  

\begin{figure}[h!]
\begin{center}
\includegraphics[width=0.45\textwidth]{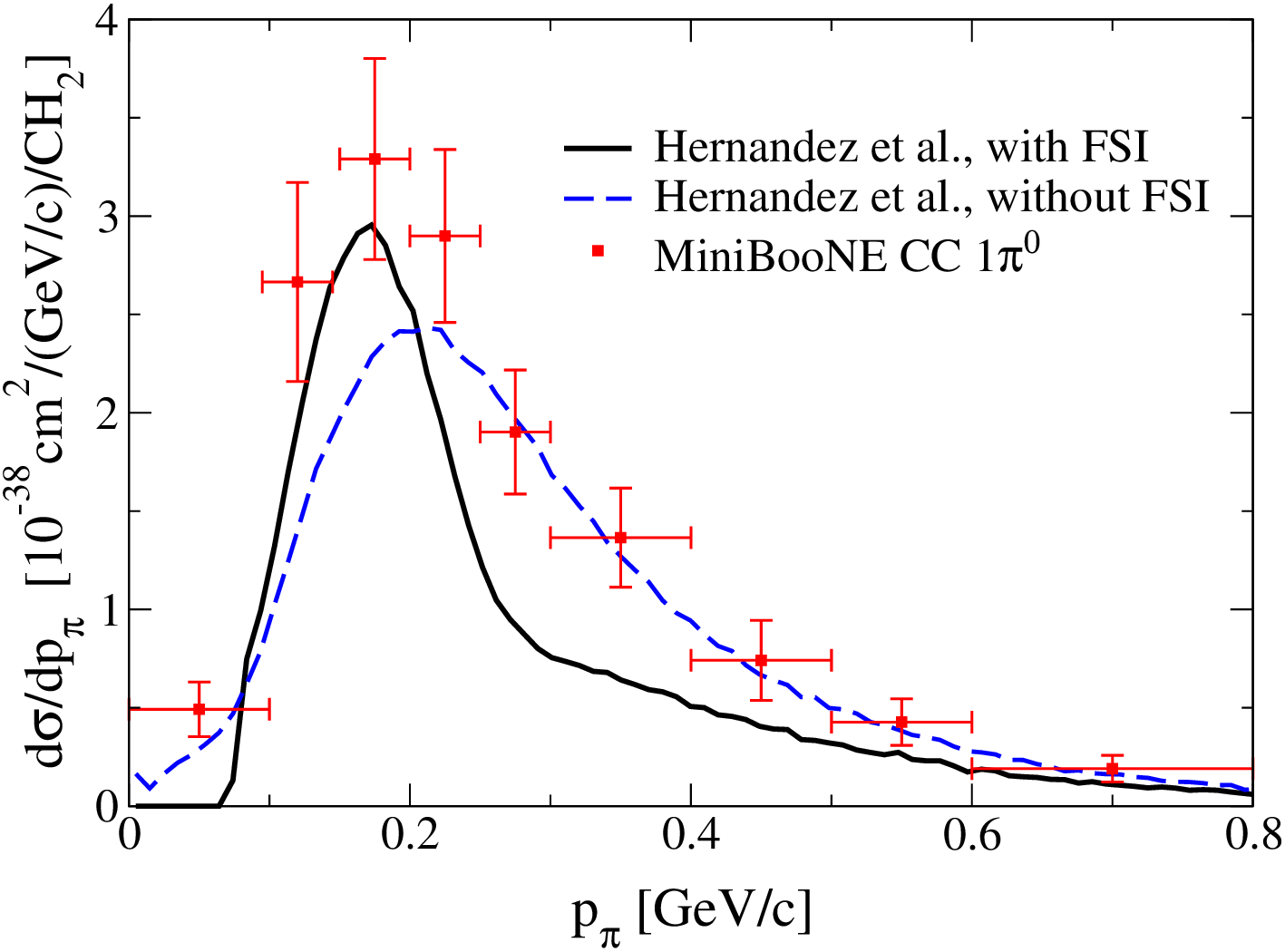}\hspace{0.05\textwidth}\includegraphics[width=0.45\textwidth]{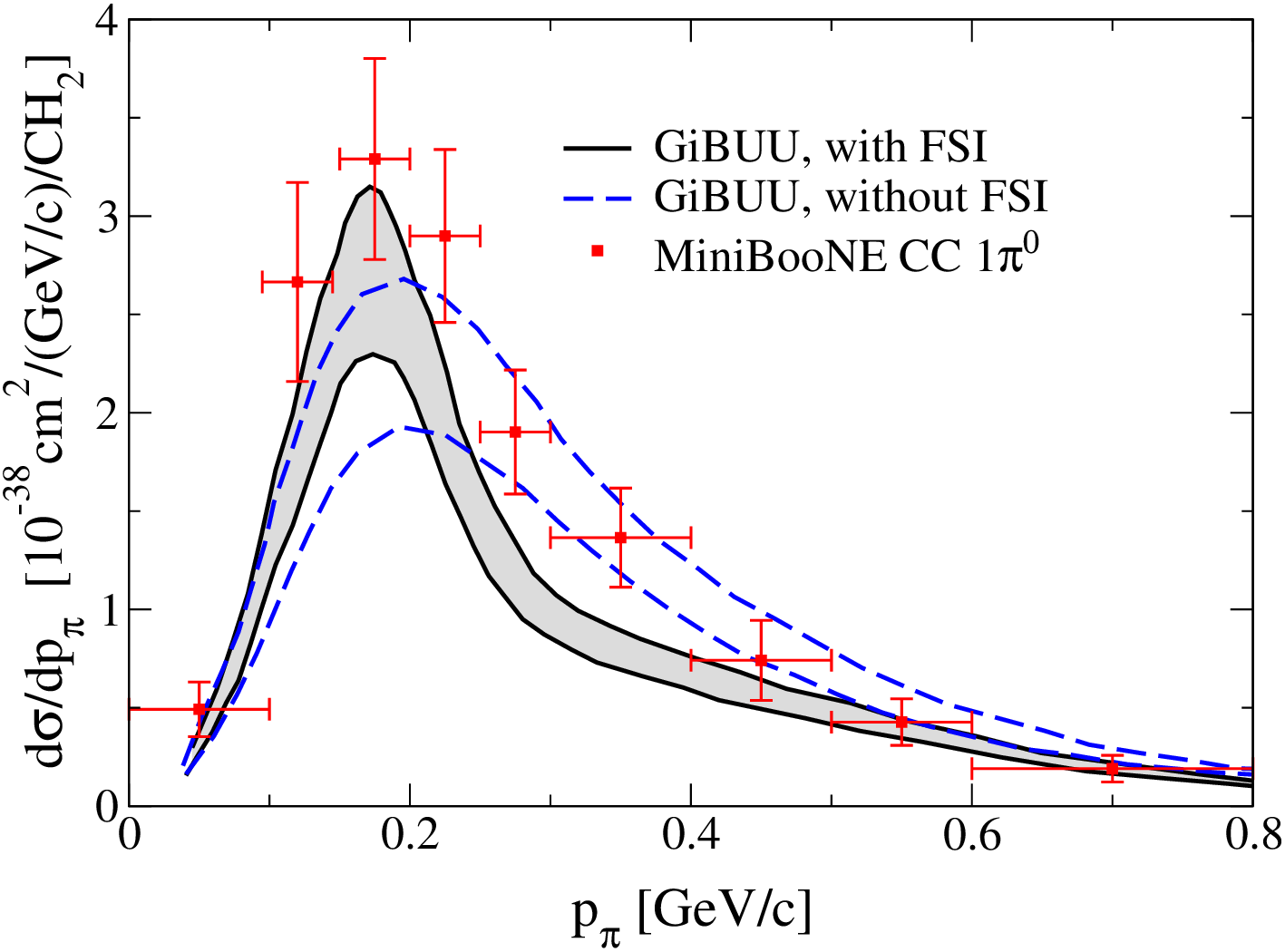}
\end{center}
\caption{CC1$\pi^0$ $d\sigma/dp_\pi$ on CH$_2$ folded with the $\nu_\mu$ flux at the MiniBooNE detector. Data are from Ref.~\cite{AguilarArevalo:2010xt}.  Left: Predictions from the cascade approach of Ref.~\cite{Hernandez:2013jka}. The solid curve corresponds to the full model and the dashed one shows the results without FSI. Right: Predictions from the GiBUU transport model~\cite{Lalakulich:2012cj}. The dashed (solid) curves were obtained without (with) FSI. Two different choices of $C_5^A(q^2)$, independently tuned to the ANL and BNL data sets give rise to the systematic uncertainty bands.}
\label{fig:CCpi0}
\end{figure}

The MiniBooNE measurements, reported as single pion momentum and angular flux-averaged distributions, have been compared to the most comprehensive approaches available~\cite{Lalakulich:2012cj,Hernandez:2013jka}. It is remarkable that in spite of the different treatment of FSI (multi-channel transport in Ref.~\cite{Lalakulich:2012cj} and pion cascade in Ref.~\cite{Hernandez:2013jka}), the two models obtain very similar results. The comparison to data, displayed in Fig.~\ref{fig:CCpi0} for CC1$\pi^0$, reveals an unexplained excess of pions with momenta between 200 and 500~MeV/c and forward angles (see also the right panels of Fig.~8 in Ref.~\cite{Hernandez:2013jka} and Fig.~9 of Ref.~\cite{Lalakulich:2012cj}). Such shrinking of the pion-momentum peak by FSI has however been observed in pion photoproduction~\cite{Krusche:2004uw}. Unaccounted multi-nucleon mechanisms could be relevant but are unlikely to explain the disagreements.

\begin{figure}[h!]
\begin{center}
\includegraphics[width=0.45\textwidth]{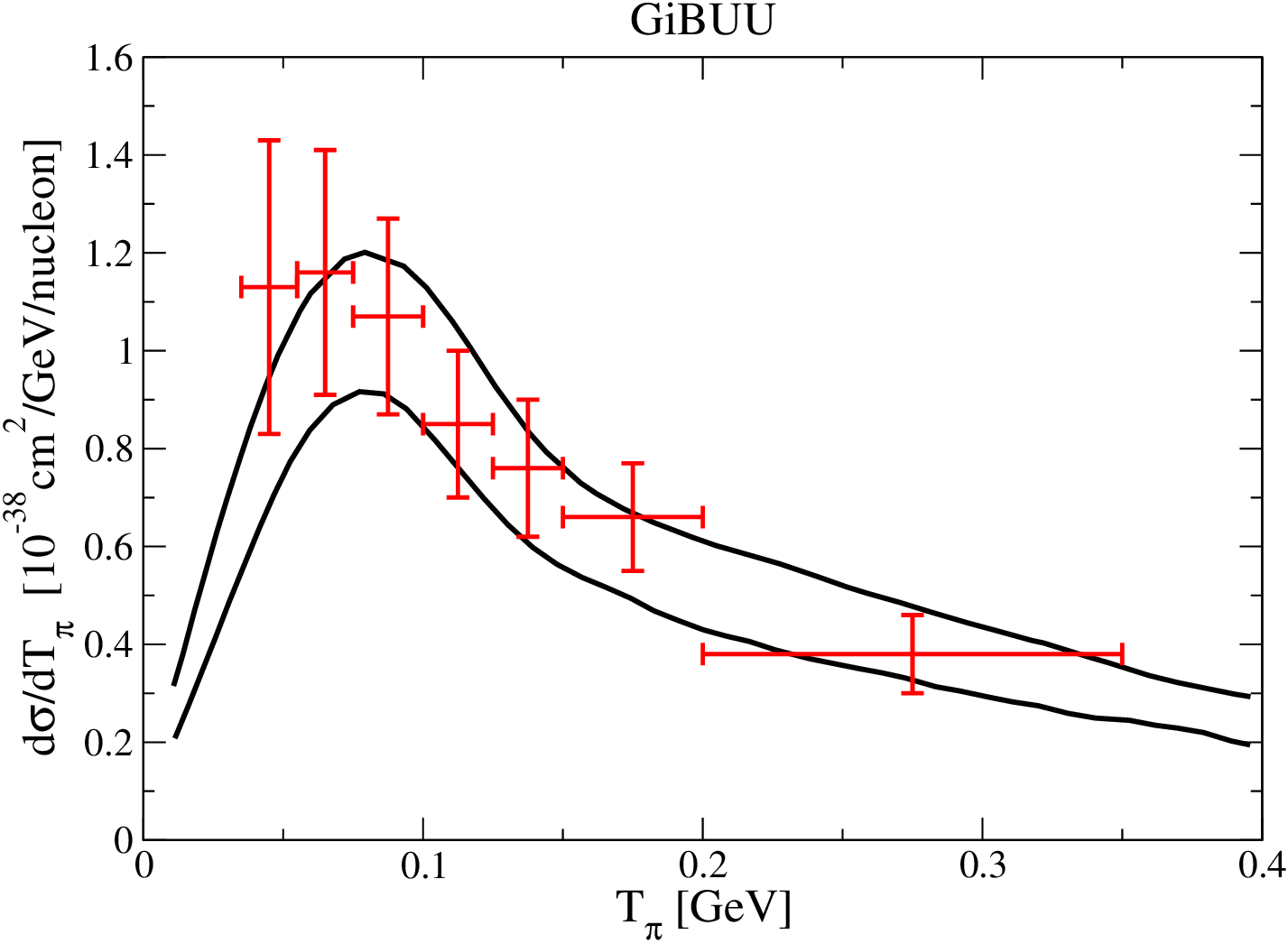}\hspace{0.05\textwidth}\includegraphics[width=0.45\textwidth]{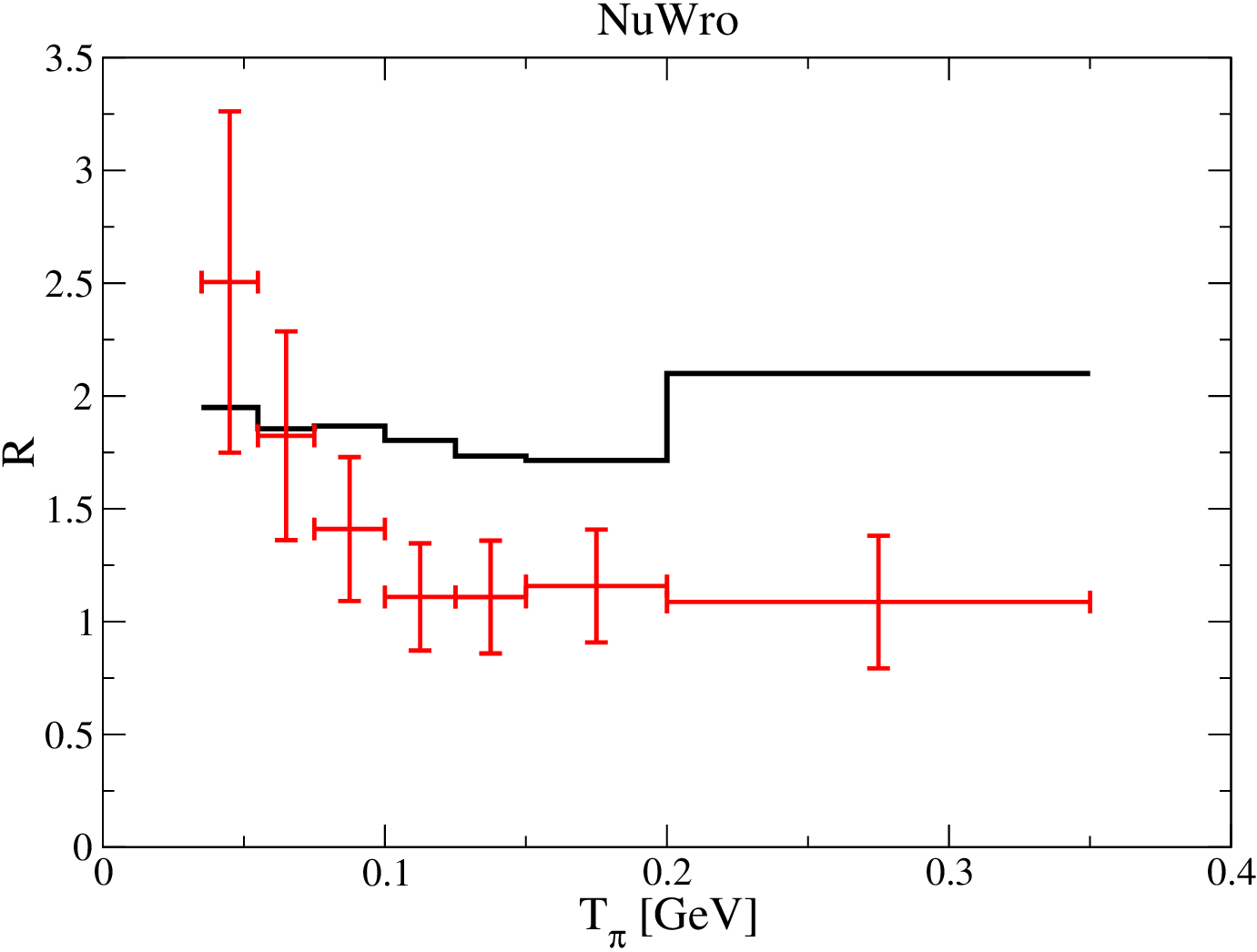}
\end{center}
\caption{Left (adapted from Ref.~\cite{Mosel:2015tja}): Differential cross section for CC$\pi^\pm$ on CH averaged over the MINERvA flux as a function of the pion kinetic energy computed with the GiBUU model. The upper (lower) curve is obtained with the nucleon cross sections tuned to BNL (ANL) data. Experimental results are from Ref.~\cite{Eberly:2014mra}. Right (adapted from Ref.~\cite{Sobczyk:2014xza}): Ratio of $d\sigma/dT_\pi$ from MiniBooNE and MINERvA [Eq.~(\ref{eq:ratio})], and the corresponding NuWro predictions.}
\label{fig:CCpi}
\end{figure}

The shape disagreement apparent in Fig~\ref{fig:CCpi0} is in contrast with the result of the GiBUU model for CC$\pi^{\pm}$ (mostly $\pi^+$) reaction compared to MINERvA data. The left panel of Fig~\ref{fig:CCpi} is adapted from Fig.~1 of Ref.~\cite{Mosel:2015tja}. The band between the two solid lines represent the uncertainty from ANL and BNL data~\cite{Mosel:2015tja}. This band would be narrower and closer to the lower end  if the reanalyzed data of Ref.~\cite{Wilkinson:2014yfa} were used. As can be seen in Fig.~2 of Ref.~\cite{Mosel:2015tja} the strength missing in the lower curve (consistent with ANL) comes from forward pion angles. 
One is tempted to attribute the different scenarios displayed by Fig~\ref{fig:CCpi0} and  Fig~\ref{fig:CCpi} (left) to the differences in the corresponding neutrino fluxes. The flux at MiniBooNE peaks at around 700~MeV while the MINERvA one does close to 3~GeV. However, according to Ref.~\cite{Sobczyk:2014xza}, there is a strong correlation among the two data sets in spite of the flux differences. Using the NuWro generators, the authors of Ref.~\cite{Sobczyk:2014xza} have obtained that the ratio
\be
R(T_\pi) =\frac{ \left(d\sigma / dT_\pi\right)_{\mathrm{MINERvA,\,CC}\pi^\pm}}
{\left(d\sigma / dT_\pi\right)_{\mathrm{MiniBooNE,\,CC}\pi^+}}  \approx 2
\label{eq:ratio}
\ee
as can be seen in the right panel of Fig.~\ref{fig:CCpi}, adapted from Ref.~\cite{Sobczyk:2014xza}. In both experiments, the dominant contribution comes from the $\Delta(1232)$ region. The cut in $W_\mathrm{rec} \equiv \sqrt{m_N^2 + 2 m_N q_0 + q^2} < 1.4$~GeV is applied in the MINERvA analysis~\cite{Eberly:2014mra} using measured lepton kinematics and calorimetry. It quenches the contribution from higher invariant masses although the cut is not sharp, and the $\Delta$ peak is shrunk from its maximum on~\cite{Mosel:2015tja}. This happens once $W_\mathrm{rec}$ does not coincide with the actual hadronic invariant mass $W^2 = (q+p)^2$ because the initial nucleon is not at rest~\cite{Wilkinson:2014yfa}. As shown in Fig.~\ref{fig:CCpi} (right), the correlation obtained in Ref.~\cite{Sobczyk:2014xza} with NuWro is absent in the data. Further progress in the understanding of weak pion production on nuclear targets requires this tension to be resolved. 

\section{Coherent pion production at MINERvA}

Thanks to the MINERvA experiment, we now have detailed information about the energy and angular distributions of pions produced in (anti)neutrino interactions on nuclei, where the target remains in the ground state~\cite{Higuera:2014azj}. A proper understanding of these coherent pion production (Coh$\pi$) data is a new challenge for model builders.

\begin{figure}[h!]
\begin{center}
\includegraphics[width=0.45\textwidth]{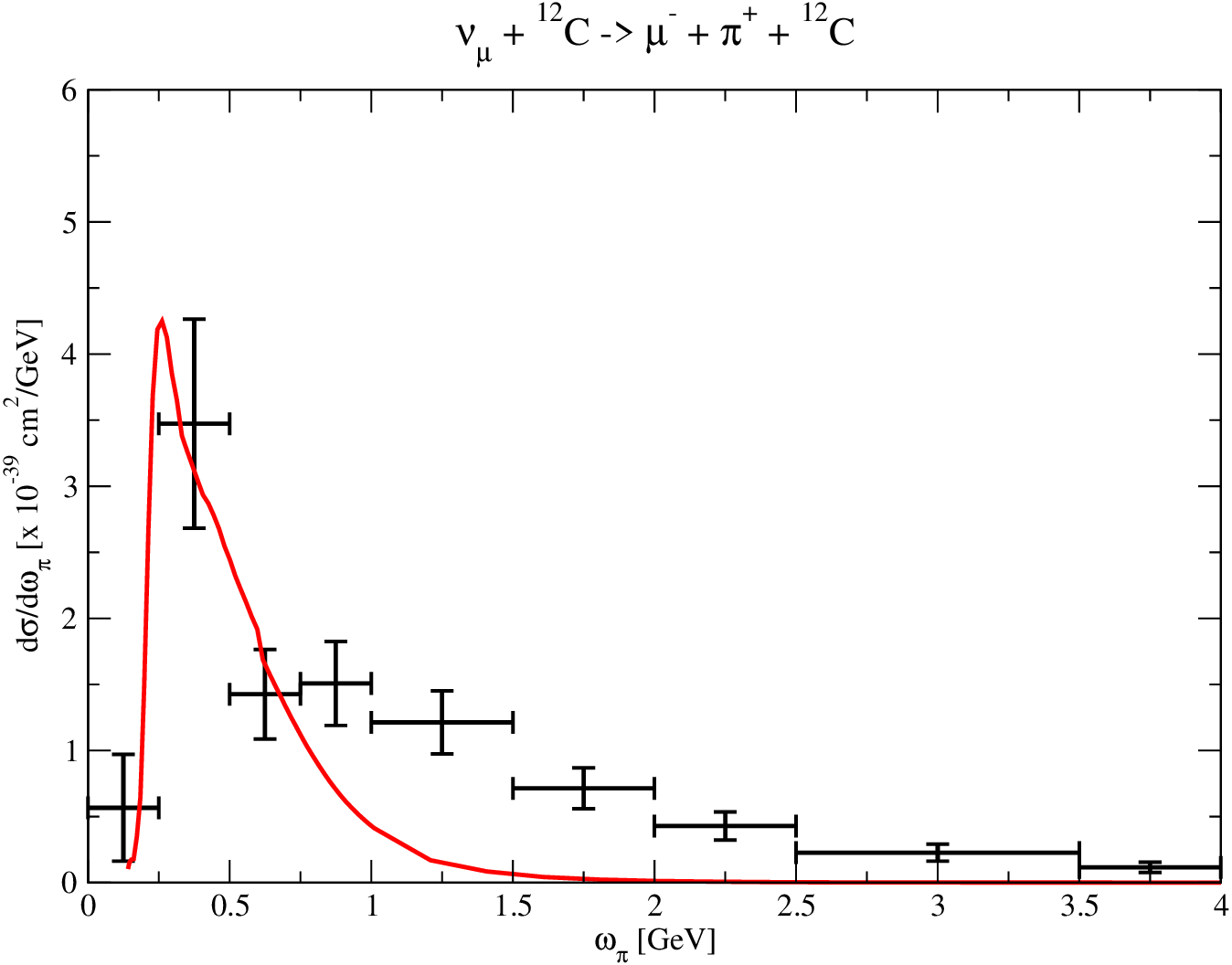}\hspace{0.05\textwidth}\includegraphics[width=0.45\textwidth]{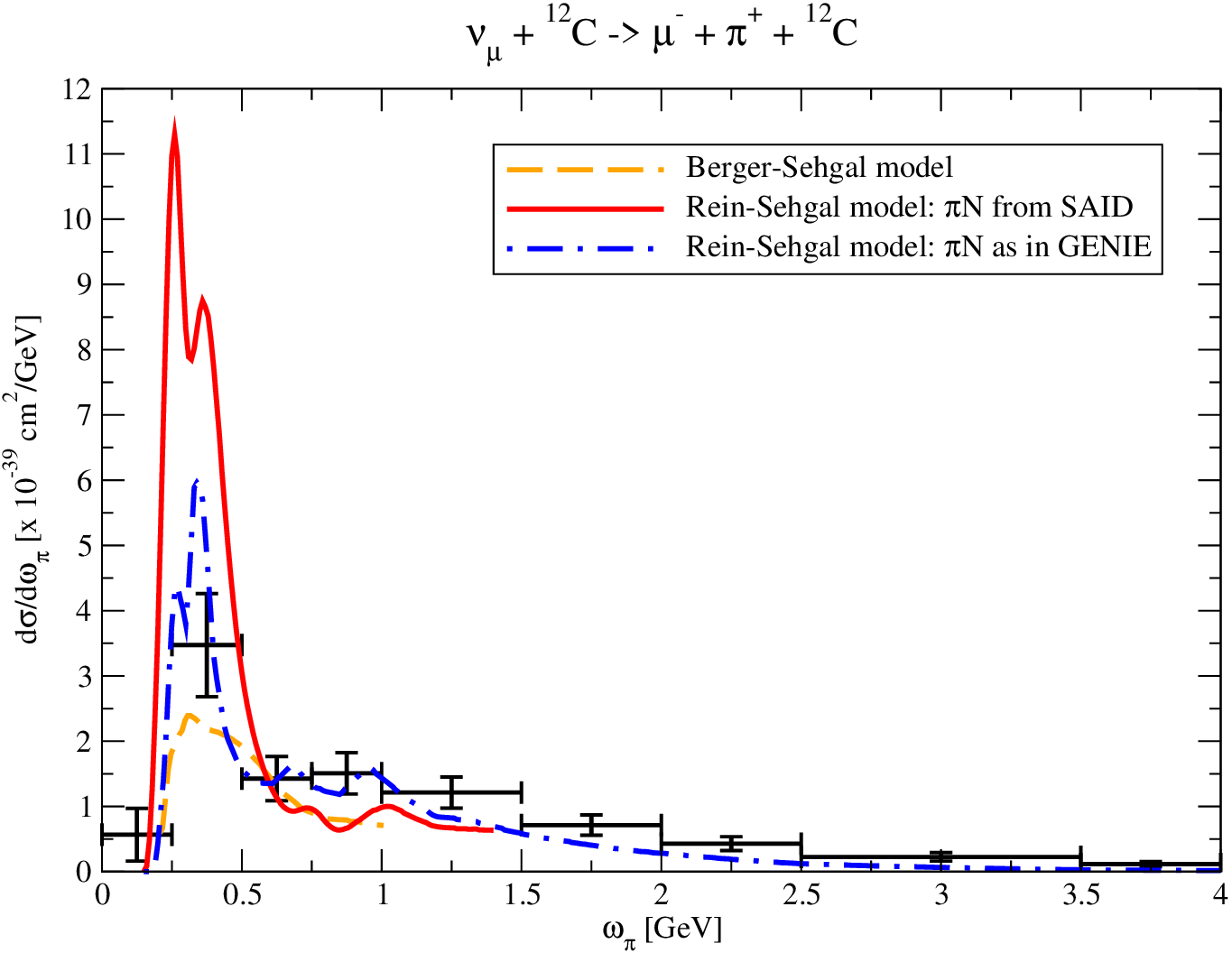}
\end{center}
\caption{Coh$\pi^+$ MINERvA data~\cite{Higuera:2014azj} confronted with different theoretical models. Left: microscopic model of Ref.~\cite{AlvarezRuso:2007tt}. Right: our implementations of Rein-Sehgal~\cite{Rein:1982pf} and Berger-Sehgal~\cite{Berger:2008xs} models. For the 
  Rein-Sehgal model, input as in GENIE~\cite{Perdue} and from the SAID analysis~\cite{SAID} have been used.}
\label{fig:CohPi}
\end{figure}

Coh$\pi$ models are traditionally classified as microscopic or PCAC ones. Microscopic approaches start with a model for pion production on the nucleon and perform a coherent sum over all nucleonic currents. Modifications of the elementary amplitudes in the nuclear medium are also
taken into account. A quantum treatment of the pion distortion is usually applied via the Klein-Gordon~\cite{AlvarezRuso:2007tt,Amaro:2008hd} or the Lippmann-Schwinger~\cite{Nakamura:2009iq} equations although the semiclassical eikonal approximation has also been employed~\cite{Singh:2006bm,Zhang:2012xi}. These models do not critically rely on PCAC although it is often present in the nucleon currents. For these reason, they can be validated with data on coherent pion photo and electroproduction. The main challenge for microscopic models developed so far is that they are restricted to the kinematic region where the excitation of the $\Delta(1232)$ is dominant. In the left panel of Fig~\ref{fig:CohPi}, the prediction of the model of Ref.~\cite{AlvarezRuso:2007tt} for the pion energy differential cross section averaged over the MINERvA flux is compared to the data of Ref.~\cite{Higuera:2014azj}. A good description is found at low pion energies, where the model is applicable, while the high energy tail is missed. Coh$\pi$ is dominated by low $q^2$. In this limit, Eq.~(\ref{eq:q2=0}) implies that the predicted cross section strongly depends on the value of the leading $N \Delta$ axial coupling $C_5^A(0)$. The results in Fig.~\ref{fig:CohPi} (left) are obtained using the GT relation. A value of $C_5^A(0)$ extracted from ANL/BNL data ignoring Watson's theorem would result in a 30\% smaller cross section.     

PCAC approaches take advantage of the fact that at $q^2=0$, Coh$\pi$ can be related to pion-nucleus elastic scattering. Based on this principle, Rein and Sehgal (RS) built a simple and elegant Coh$\pi^0$ model using empirical information about pion-nucleon elastic and inelastic scattering~\cite{Rein:1982pf}. A common issue of PCAC models is that the $q^2 =0$ approximation neglects terms in the cross section that vanish in this limit but not at finite $q^2$, leading to pion angular distributions that are too wide~\cite{Hernandez:2009vm}. But the main problem of the RS model resides on its poor description of pion-nucleus elastic scattering (see Fig.~2 of Ref.~\cite{Hernandez:2009vm}). This was improved in Refs.~\cite{Berger:2008xs,Paschos:2009ag} by the direct use of experimental pion-nucleus elastic cross sections although, in this way, the off-shell dependence of the pion-nucleus amplitude due to the fact that $q^2 \lesssim 0$ probed in Coh$\pi$ is different from $m_\pi^2$ for real pions is neglected. In Fig.~\ref{fig:CohPi} (right) we compare our own implementation of the RS~\cite{Rein:1982pf} and the Berger-Sehgal (BS)~\cite{Berger:2008xs} approaches to the MINERvA data. Within the RS model we consider the $\pi N$ parametrizations as implemented in GENIE~\cite{Perdue} as well as the state-of-the-art ones from SAID~\cite{SAID}. The plot shows that the RS cross section is very sensitive to this input. An improvement in the parametrizations does actually cause a worse agreement with data. From this perspective, the good agreement obtained by the GENIE implementation, particularly above $\omega_\pi = 500$~MeV, (see also Fig.~4 of Ref.~\cite{Higuera:2014azj} can be regarded as accidental. The prediction from the BS model is better but not entirely satisfactory as it underestimates both the low-energy peak and the region of  $\omega_\pi = 0.6 - 1$~GeV.

\section{Summary}
We have discussed some of the recent developments in the physics of weak pion production from a theoretical perspective. The importance of respecting Watson's theorem in order to reconcile ANL/BNL data with the off-diagonal GT relation is stressed. It is shown that the reanalyzed and consistent ANL and BNL data result in a more precise determination of the leading axial coupling $C_5^A(0)$ but do not constrain better other parameters in the axial current. The disagreement of pion production calculations on nuclei with MiniBooNE data is presented in the light of the new MINERvA data. The absence of the strong correlations between the two data sets, in disagreement with the NuWro simulation, as found by Sobczyk and Zmuda, needs to be understood. It is shown that a microscopic Coh$\pi$ model is able to describe the low-energy peak in the MINERvA data, which is dominated by $\Delta(1232)$ excitation, but fails at higher energies. The difficulties of PCAC models to reproduce the experiment and, in particular, the sensitivity of the RS model to the $\pi N$ input are also presented.

\section*{Acknowledgements}
Research supported by the Spanish Ministerio de Econom\'\i a y Competitividad and the European FEDER funds, under Contracts FPA2013-47443-C2-2-P, FIS2014-51948-C2-1-P, FIS2014-51948-C2-2-P,  FIS2014-57026-REDT and SEV-2014-0398, by Generalitat Valenciana under Contract PROMETEOII/2014/0068 and by the European Union HadronPhysics3 project, grant agreement no. 283286.

\bibliography{neutrinos}

\end{document}